\documentclass[epj]{svjour}
\usepackage{amsfonts,amssymb}
\usepackage{amsmath}

\begin{document}

\title{Canonical Analysis of Scalar Fields in Two Dimensional Curved Space}

\abstract{
Scalar fields on a two dimensional curved surface are considered and the canonical structure of this theory analyzed.  Both the first and second order forms of the Einstein-Hilbert (EH) action for the 
metric are used (these being inequivalent in two dimensions).  The Dirac constraint formalism is used to find the generator 
of the gauge transformation, using the formalisms of Henneaux, Teitelboim and Zanelli (HTZ) and of Castellani (C).  The HTZ formalism is slightly 
modified in the case of the first order EH action to accommodate the gauge transformation of the metric; this gauge transformation is unusual as it mixes the affine connection 
with the scalar field.}


\author{D.G.C. McKeon$^{1,2,a}$  \and Alexander Patrushev$^{1,b}$} 

\institute{Department of Applied Mathematics, The
University of Western Ontario, London, ON N6A 5B7, Canada \and Department of Mathematics and
Computer Science, Algoma University, Sault St.Marie, ON P6A
2G4, Canada}

\date{}

\maketitle
\email{$^a$dgmckeo2@uwo.ca, $^b$apatrush@uwo.ca}

\section{Introduction}
Scalar fields on a curved background have received considerable attention because of their relationship with Bosonic string theory \cite{21}. One normally focuses on the quantum properties of string theory 
(such as the absence of the conformal anomaly only if the dimension of the target space exceeds four), but it is both interesting and important to have an understanding of the classical canonical structure
of this model if one is to truly comprehend the implications of the quantization procedure. In this paper we undertake the task of applying Dirac's analysis of constrained systems \cite{1,2,3,4,5,6} 
to the problem of $N$ scalar fields on a curved two dimensional manifold. We focus in particular on the first class constraints that appear and what they tell us about the gauge invariance present in the theory.
A number of novel features arise.
 
  Generally, in any discussion of metric fields on a two dimensional space, the action for the metric is ignored as the Einstein-Hilbert (EH) action $\sqrt{-g}g^{\mu\nu}R_{\mu\nu}(g_{\alpha\beta})$ in two
dimensions (2D), when treated as a function of the metric $g_{\mu\nu}$ alone (the second order form), is a pure surface term and has no dynamical degrees of freedom. We note though that this lack of dynamics
does not mean that it cannot be quantized; this has been studied in refs. \cite{30,31} using a BRST analysis. There has also been a discussion of the canonical structure of the first order EH action 
in 2D \cite{7}. The first class constraints that occur have been shown to imply that there is an an invariance under the gauge transformation
\begin{equation}
g_{\mu \nu} \rightarrow g_{\mu \nu} +\omega_{\mu\nu}
\end{equation}
which is consistent with there being no degrees of freedom present in the action.
Normally when a matter field is coupled with a gauge field (eg. the electron is coupled to a photon), any gauge invariance present in the uncoupled gauge field action is respected by the action in which 
the coupling is present. In this case however, the coupling of $N$ scalars $f^a, (a=1,2...N)$ to the metric $g_{\mu\nu}$ through the Lagrangian
\begin{equation}
\mathcal{L}_f = \frac{1}{2} \sqrt{-g}\, g^{\mu\nu} \partial_\mu f^a \partial_\nu f^a,
\end{equation}  
while being diffeomorphism invariant, does not respect the symmetry of eq. (1). In this paper, we first address the problem of disentangling how supplementing the second order order
EH action in 2D by the action of eq. (2) alters the constraint structure of the theory and thereby leads to a new gauge 
invariance that is distinct from that of eq. (1).

 The problem of reconciling the gauge invariance present in the action for the free gauge field with that occurring when it is coupled to a matter field becomes even more interesting when the free gauge action 
is the first order EH action in 2D. We first note that this action, $\sqrt{-g}g^{\mu\nu}R_{\mu\nu}(\Gamma^{\lambda}_{\alpha\beta}),$ is not equivalent to the second order form, unlike what occurs in $D>2$ 
dimensions \cite{14,15}. This is because the affine connection $\Gamma^{\lambda}_{\mu\nu}$ is no longer given by the Christoffel symbol
\begin{equation}
\left\lbrace \begin{array}{c}
\lambda \\
\mu\nu
\end{array} \right\rbrace = \frac{1}{2} g^{\lambda\sigma} \left(g_{\sigma\mu ,\nu} + g_{\sigma\nu ,\mu} - g_{\mu\nu , \sigma}\right)
\end{equation}
but rather
\begin{equation}
\Gamma_{\mu\nu}^\lambda = 
\left\lbrace \begin{array}{c}
\lambda \\
\mu\nu
\end{array} \right\rbrace + \delta_\mu^\lambda\xi_\nu + \delta_\nu^\lambda\xi_\mu -
g_{\mu\nu} \xi^\lambda
\end{equation}
(where $\xi^\lambda$ is an arbitrary vector) when solving the equation of motion for $\Gamma_{\mu\nu}^\lambda$. We first consider the implication of having this extra field arising in the model. We then review
analysis \cite{8,9,10,11,12,13} which shows that the canonical structure of the first order EH action in 2D shows that there are no physical degrees of freedom in the model despite it not being topological, 
and that the first class constraint that arise result in a novel gauge transformation 
\begin{equation}
\delta h^{\mu\nu} = - \left(\epsilon^{\mu\rho} h^{\nu\sigma} + \epsilon^{\nu\rho}h^{\mu\sigma}\right)\omega_{\rho\sigma}
\end{equation}
\begin{equation}
\delta G_{\mu\nu}^\lambda = - \epsilon^{\lambda\rho} \omega_{\mu\nu ,\rho} - \epsilon^{\rho\sigma}
\left(G_{\mu\rho}^\lambda \omega_{\nu\sigma} + G_{\nu\rho}^\lambda \omega_{\mu\sigma}\right)
\end{equation}
where $\epsilon^{01} = -\epsilon^{10} = 1$, $\epsilon^{00} = \epsilon^{11} = 0$, $h^{\mu\nu} = \sqrt{-g}\, g^{\mu\nu}$ 
and $G_{\mu\nu}^\lambda = \Gamma_{\mu\nu}^\lambda - \frac{1}{2} \left(\delta_\mu^\lambda \Gamma_{\rho\nu}^\rho + \delta_\nu^\lambda \Gamma_{\rho\mu}^\rho\right)$.  
This is distinct from the manifest diffeomorphism invariance present. We then address the problem of seeing how the first class constraints that lead to eqs. (5,6) are modified when the free action for 
$h^{\mu\nu},G_{\mu\nu}^\lambda$ is supplemented by
\begin{equation}
\mathcal{L}_f = \frac{1}{2} h^{\mu\nu} \partial_\mu f^a\,\partial_\nu f^a.
\end{equation} 
 
 A number of interesting features arise in the course of applying the Dirac constraint formalism to these two models in which a scalar field propagates on a curved surface. First of all, when there are $N$
scalar fields, the constraints and their associated gauge conditions combine to leave just $2N-4$ dynamical degrees of freedom in the theory. Secondly, when one considers either the first or second order EH
action to be the action for the gauge field coupled to the scalar matter field, the number of first class constraints in each generation is not the same. For $N=1$, there are in the case of the second order EH 
action, three primary and two secondary first class constraints, while with the first order EH action there are three primary and secondary first class constraints and two tertiary first class constraints.
Consequently, when using these constraints to find the gauge invariance that they imply to be present in the initial action, one finds that the techniques of both C (refs. \cite{16,17}) and of HTZ
(refs. \cite{18,19}) do not lead to a unique gauge transformation. Neither diffeomorphism invariance not conformal invariance are implied by these first class constraints; indeed for the first order action 
the gauge generator derived from the first class constraints implies that the scalar field and affine connections mix under a gauge transformations.
 
 In the next two sections we present a canonical analysis of a scalar field on a curved background, using the second, then the first, order EH action for the metric, including a discussion of the gauge 
transformations implied by the first class constraints. In appendix, the way in which the first class constraints can be used to find the generator of the gauge transformation is outlined, using both the approach
of C \cite{16,17} and of HTZ \cite{18,19}.

\section{Second order EH Action and Scalar Fields}

We begin by first reviewing how the second order EH action in 2D can be treated using the Dirac constraint formalism \cite{7}, despite it being a topological theory.
We then couple the metric to a scalar field and consider how this affects the gauge invariance of eq. (3).

The second order EH action is
\begin{equation}
S_{EH} = \int dx \sqrt{-g}\, g^{\mu\nu} R_{\mu\nu}
\end{equation}
where
\begin{equation}
R_{\mu\nu} = \Gamma^\lambda_{\mu\nu ,\lambda} - \Gamma^\lambda_{\lambda\mu ,\nu} + \Gamma_{\lambda\sigma}^\lambda \Gamma_{\mu\nu}^\sigma - 
\Gamma_{\sigma\mu}^\lambda \Gamma_{\lambda\nu}^\sigma
\end{equation}
and $\Gamma_{\mu\nu}^\lambda = \left\lbrace \begin{array}{c}
\lambda \\
\mu\nu \end{array}\right\rbrace$.  
In any dimension \cite{20}
\begin{align}
\sqrt{-g}& \, g^{\mu\nu} \left(\Gamma_{\mu\nu ,\lambda}^\lambda - \Gamma_{\lambda\mu , \nu}^\lambda \right)\nonumber \\ 
&= \left(\sqrt{-g}\, g^{\mu\nu} \Gamma_{\mu\nu}^\lambda\right)_{,\lambda} - 
\left(\sqrt{-g}\, g^{\mu\nu} \Gamma_{\lambda\mu}^\lambda\right)_{,\nu} \nonumber \\
&-2\sqrt{-g}\, g^{\mu\nu} \left(\Gamma_{\lambda\sigma}^\lambda \Gamma_{\mu\nu}^\sigma - \Gamma_{\sigma\mu}^\lambda \Gamma_{\lambda\nu}^\sigma\right)
\end{align}
and hence if surface terms are discarded, then $S_{EH}$ can be replaced by the non-covariant action
\begin{equation}
S_{\Gamma\Gamma}^{(2)} = - \int dx \sqrt{-g}\, g^{\mu\nu} \left(\Gamma_{\lambda\sigma}^\lambda \Gamma_{\mu\nu}^\sigma - \Gamma_{\sigma\mu}^\lambda \Gamma_{\lambda\nu}^\sigma\right). 
\end{equation}  
It is this form of the action that was used by Dirac in the analysis of the canonical structure of the EH action in 4D \cite{22}. (See also refs \cite{32,33}.) 
We too will use it as the initial action for analyzing the EH action in 2D. 

In 2D, eq. (11) becomes 
\begin{align}
= \frac{1}{2} \int dx &(-g)^{-3/2} \left[ g_{11,0}\left(g_{01}g_{00,1} - g_{00} g_{01,1}\right)\right. \\
& + g_{00,0} \left(g_{11} g_{01,1} - g_{01}g_{11,1}\right) \nonumber \\
& \left. + g_{01,0} \left(g_{00} g_{11,1} - g_{11}g_{00,1}\right)\right]. \nonumber
\end{align}
If one were to choose conformal coordinates so that $g_{00} = -g_{11} = \rho(x)$, $g_{01} = 0$ as in \cite{21}, then $S_{\Gamma\Gamma}$ vanishes.  However, if $g_{01} \neq 0$ then 
$S_{\Gamma\Gamma}$ is amenable to canonical analysis \cite{7}.  However, it becomes apparent that $S_{\Gamma\Gamma}$ itself is a surface term if we adopt the coordinates \cite{22}
\begin{equation}
\delta = \frac{-\sqrt{-g}}{g_{11}}\;\;\;, \rho = \frac{g_{01}}{g_{11}}\;\;\;\;, g_{11}
\end{equation}
so that
\begin{align}
S_{\Gamma\Gamma}^{(2)} &= \int dx \frac{1}{\delta^2} \left(\delta_{,0}\rho_{,1} - \rho_{,0}\delta_{,1}\right)\nonumber \\
&= \int dx \left[\left(\frac{\rho_{,0}}{\delta}\right)_{,1} - \left(\frac{\rho_{,1}}{\delta}\right)_{,0} \right].
\end{align}

We will not employ the variables $\delta$ and $\rho$ in our canonical analysis; they simply serve to simplify the demonstration that $S^{(2)}_{\Gamma\Gamma}$ is a surface term. They do appear in ref. \cite{25} though.

From eq. (12), we find the primary constraints 
\begin{subequations}
    \begin{eqnarray}
\chi^{11}&=& \pi^{11} - \frac{1}{2(-g)^{3/2}} \left(g_{01} g_{00,1} - 
g_{00} g_{01,1}\right)\\
\chi^{00}&=& \pi^{00} - \frac{1}{2(-g)^{3/2}} \left(g_{11} g_{01,1} - 
g_{01} g_{11,1}\right)\\
\chi^{01}&=& \pi^{01} - \frac{1}{2(-g)^{3/2}} \left(g_{00} g_{11,1} - 
g_{11} g_{01,1}\right)
\end{eqnarray}
\end{subequations}
where $(\pi^{11}, \pi^{00}, \pi^{01})$ are the canonical momenta conjugate to $(g_{11}, g_{00}, g_{01})$ respectively. (If one were to simply discard the action of eq.(12) 
because of its topological nature, 
then we would merely have $\chi^{11}=\pi^{11},\,\chi^{00}=\pi^{00}$ and $\chi^{01}=\pi^{01}$.) The Poisson Bracket (PB) of any two of these 
constraints vanishes.  Furthermore, the canonical Hamiltonian vanishes.  Consequently there are three primary first class constraints and no secondary constraints associated 
with $S_{\Gamma\Gamma}^{(2)}$ using any of the techniques of refs. \cite{16,17,18,19} one finds the generator of gauge transformations to be
\begin{equation}
G = \int dx \left[ \omega_{11} \chi^{11} + \omega_{00}\chi^{00} + \omega_{01}\chi^{01}\right]
\end{equation}
which results in 
\begin{equation}
\delta g_{\mu\nu} = \omega_{\mu\nu}
\end{equation}
as in eq. (3). Eq. (17) also would follow from just taking $\chi^{11}=\pi^{11},\,\chi^{00}=\pi^{00}$ and $\chi^{01}=\pi^{01}$, as is appropriate if were to discard
the action all together because of it being topological.

We note that with these first class constraints of eq. (15) and the three associated gauge conditions,
these are six restrictions on the six canonical variables ($g_{\mu\nu}$ and $\pi^{\mu\nu}$) in phase space, leaving no physical degrees of freedom.
Supplementing $S_{\Gamma\Gamma}^{(2)}$ with the action for a massless scalar field $f$ \cite{23}
\begin{equation}
S_f = \frac{1}{2} \int dx \sqrt{-g} \,g^{\mu\nu} f_{,\mu}f_{,\nu}
\end{equation}
we find that the momentum conjugate to $f$ is
\begin{equation}
p = \sqrt{-g} \left(g^{00} f_{,0} + g^{01}f_{,1}\right) = \frac{1}{\sqrt{-g}} \left(-g_{11} f_{,0} + g_{01}f_{,1}\right)
\end{equation}
so that the part of the canonical Hamiltonian arising from $S_f$ in eq. (18) is 
\begin{equation}
\mathcal{H}_c = \delta S+\rho I\!\!P
\end{equation}
where $S$ and $I\!\!P$ are two new secondary constraints
\begin{align}
S &= \frac{1}{2}\left(p^2 + f_{,1}^2\right)\\
I\!\!P &= pf_{,1} .
\end{align}
We note that although only the combinations $\delta$ and $\rho$ enter both eqs. (14) and (20), all three components of $h^{\mu\nu}$ appear in
the initial action of eqs. (11) and (18). These three must be all included be as fields in the canonical analysis. In ref. \cite{21}, a special "conformal gauge''
was used to dispense with the "conformal factor'' contribution to the action of eq. (18), reducing the number of independent components of the metric from three to two.
However, choosing a "gauge'' at the outset of any canonical analysis is inconsistent with Dirac's procedure \cite{1,2,3,4,5}.

Using test functions as in ref. \cite{24} we find 
\begin{subequations}
    \begin{eqnarray}
\left\lbrace S(x), S(y) \right\rbrace &=&  \left( - I\!\!P (x) \partial^y_1 + I\!\!P (y)\partial^x_1\right) 
\delta (x-y) \nonumber \\
&=& \left\lbrace I\!\!P(x), I\!\!P(y)\right\rbrace \\
\left\lbrace I\!\!P(x), S(y) \right\rbrace &=& \left( - S(x) \partial^y_1 + S(y)\partial^x_1\right) 
\delta (x-y) \nonumber \\
&=& \left\lbrace S(x), I\!\!P(y)\right\rbrace
\end{eqnarray}
\end{subequations}
and thus no tertiary constraints arise.

With eqs. (15,21,22) we see that there are now five first class constraints, which when combined with five associated gauge conditions,
leaves us with ten restrictions on the eight variables in phase space $g_{\mu\nu},\,f$ and their associated momenta).
If the single scalar field $f$ in eq.(6) were replaced by $N$ scalars $f^a~~(a=1,2...N)$ in an $O(N)$ symmetric fashion, 
there still would be ten constraints in phase space, but there would now be $2N+6$ variables, leaving $2N-4$ net physical degrees of freedom. 
Only if $N>2$ are there true physical degrees of freedom.

The general form of the gauge generator for $S_{\Gamma\Gamma}^{(2)} + S_f$, when using the HTZ approach \cite{18,19}, is 
\begin{align}
G_{HTZ} = \int dx(A_{11} \chi^{11} &+ A_{00} \chi^{00} + A_{01} \chi^{01}\nonumber \\
                        &+ B_S S + B_{I\!\!P} I\!\!P)
\end{align}
with $(A_{11}, A_{00}, A_{01})$ being found in terms of $B_S$ and $B_{I\!\!P}$ by using eq. (A5).  (In ref. \cite{25} no 
consistent way of deriving the generator of gauge transformations was used; its form is merely postulated.)

Together, eqs. (15, 20, 23) lead to eq. (A5) being satisfied to order $S$ and $I\!\!P$ provided
\begin{subequations}
    \begin{eqnarray}
(B_{I\!\!P})_{,0} &+& B_S \left( -\frac{\sqrt{-g}}{g_{11}}\right)_{,1} - (B_S)_{,1} \left(- \frac{\sqrt{-g}}{g_{11}}\right) \nonumber \\
&+& B_{I\!\!P}\left( \frac{g_{01}}{g_{11}}\right)_{,1}
-(B_p)_{,1} \left( \frac{g_{01}}{g_{11}}\right)  \nonumber \\
&+& \frac{1}{g_{11}}\left( \frac{g_{01}}{g_{11}} A_{11} - A_{01}\right) = 0
\end{eqnarray}
and \nonumber \\
\begin{eqnarray}
(B_S)_{,0} &+& B_S \left( \frac{g_{01}}{g_{11}}\right)_{,1} - (B_S)_{,1} \frac{g_{01}}{g_{11}} \\
&+&  B_{I\!\!P} \left( -\frac{\sqrt{-g}}{g_{11}}\right)_{,1}-(B_{I\!\!P})_{,1} \left( -\frac{\sqrt{-g}}{g_{11}}\right)  \nonumber \\
&+& \left[- \frac{\sqrt{-g}}{g_{11}^2} - \frac{g_{00}}{2g_{11}\sqrt{-g}} \right]A_{11} \nonumber \\ 
\qquad &-& \frac{1}{2\sqrt{-g}} A_{00} + \frac{g_{01}}{g_{11}\sqrt{-g}} A_{01} =0. \nonumber
\end{eqnarray}
\end{subequations}
As there are only two secondary constraints following from three primary constraints, eq. (25) does not uniquely fix $A_{00}$, $A_{11}$ and $A_{01}$ in terms of 
$B_S$ and $B_p$.

In any case, eq. (25) is difficult to deal with, so we will employ the approach of C which involves equations of the form of eq. (A12). In this approach, the 
form of the primary constraints that are used affects the form of the gauge generator \cite{26}.  We find it most convenient to use as primary constraints 
expressions suggested by the momenta conjugate to $\rho$, $\delta$ and $g_{11}$ under a canonical transformation:
\begin{subequations}
\begin{eqnarray}
\overline{\chi}^\rho &=& 2\chi^{00} g_{01} + \chi^{01} g_{11}\\
\overline{\chi}^\delta &=& 2\chi^{00} \sqrt{-g} \\
\overline{\chi}^{11} &=& \chi^{11} + \chi^{00} \left(\frac{ g_{00}}{g_{11}}\right) + \chi^{01} \left(\frac{ g_{01}}{g_{11}}\right)
\end{eqnarray}
\end{subequations}
so that
\begin{equation}
\left\lbrace \overline{\chi}^\rho, \mathcal{H}_c \right\rbrace = -I\!\!P,\;\;\;
\left\lbrace \overline{\chi}^\delta , \mathcal{H}_c \right\rbrace = -S,\;\;\;
\left\lbrace \overline{\chi}^{11}, \mathcal{H}_c \right\rbrace = 0.
\end{equation}

In eq. (A12), derived by using the approach of C \cite{16,17}, we take 
\begin{equation}
G_1^\rho = \overline{\chi}^\rho
\end{equation}
so that
\begin{equation}
G_0^\rho + \left\lbrace G_1^\rho , H_T\right\rbrace = \rm{p.c.}\nonumber
\end{equation}
which leads to 
\begin{align}
G_0^\rho(x) &= I\!\!P (x) + \int dy \left[\alpha_{\rho\rho} (x-y) \overline{\chi}^\rho (y)\right. \nonumber \\
 &+ \left. \alpha_{\rho\delta} (x-y) \overline{\chi}^\delta (y) + \alpha_{\rho 11} (x-y) \overline{\chi}^{11}(y)\right].
\end{align}
In turn, we must now have by eq. (A12)
\begin{equation}
\left\lbrace G_0^\rho , H_T \right\rbrace = \rm{p.c.}
\end{equation}
which fixes
\begin{align}
\int dx & \epsilon^\rho (x) G_0^\rho (x)  \nonumber \\
&= \int dx  \left[ \epsilon^\rho I\!\!P + \overline{\chi}^\rho \left(
\epsilon_{,1}^\rho \left(\frac{g_{01}}{g_{11}}\right) - \epsilon^\rho \left(\frac{g_{01}}{g_{11}}\right)_{,1}\right)\right. \nonumber \\
&+ \left. \overline{\chi}^\delta \left(\epsilon_{,1}^\rho 
\left(\frac{-\sqrt{-g}}{g_{11}}\right) - \epsilon^\rho \left(\frac{-\sqrt{-g}}{g_{11}}\right)_{,1}\right)\right].
\end{align}

So also, if 
\begin{equation}
G_1^\delta = \overline{\chi}^\delta
\end{equation}
then eq. (A12) leads to 
\begin{align}
&\int dx \epsilon^\delta (x) G_0^\delta (x) = \int dx \left[ \epsilon^\delta S + \overline{\chi}^\delta \left( \epsilon_{,1}^\delta \left(
\frac{g_{01}}{g_{11}}\right)\right.\right.\\
&- \left.\left. \epsilon^\delta \left(\frac{g_{01}}{g_{11}}\right)_{,1}\right) + \overline{\chi}^\rho 
\left(\epsilon_{,1}^\delta \left(\frac{-\sqrt{-g}}{g_{11}}\right) - \epsilon^\delta \left(\frac{-\sqrt{-g}}{g_{11}}\right)_{,1}\right)\right];\nonumber
\end{align}
we finally obtain the full generator
\begin{align}
&G_C = \int dx \left\{\epsilon^\rho I\!\!P + \epsilon^\delta S + \epsilon^{11} \overline{\chi}^{11}\right.\\
&+ \overline{\chi}^\rho \left(
\epsilon_{,1}^\rho \left(\frac{g_{01}}{g_{11}}\right) - \epsilon^\rho 
\left(\frac{g_{01}}{g_{11}}\right)_{,1} \right. \nonumber \\
&\quad + \left. \epsilon_{,1}^\delta 
\left(\frac{-\sqrt{-g}}{g_{11}}\right) - \epsilon^\delta \left(\frac{-\sqrt{-g}}{g_{11}}\right)_{,1}\right)\nonumber \\
&+ \overline{\chi}^\delta \left(
\epsilon_{,1}^\delta \left(\frac{g_{01}}{g_{11}}\right) - \epsilon^\delta 
\left(\frac{g_{01}}{g_{11}}\right)_{,1}\right. \nonumber \\
&\quad +\left. \epsilon_{,1}^\rho 
\left(\frac{-\sqrt{-g}}{g_{11}}\right) - \epsilon^\rho \left(\frac{-\sqrt{-g}}{g_{11}}\right)_{,1}\right)\nonumber \\
&+ \left. \dot{\epsilon}^\rho \overline{\chi}^\rho + \dot{\epsilon}^\delta \overline{\chi}^\delta\right\} \nonumber
\end{align}
by eq. (A10).

A third approach is to find the gauge generator, again using the HTZ approach of eq. (A5), but this time 
employing the primary constraints of eq. (26) so that 
\begin{equation}
\overline{G}_{HTZ} = \int dx \left( \overline{A}_\rho\overline{\chi}^p + \overline{A}_\delta\overline{\chi}^\delta + 
\overline{A}_{11}\overline{\chi}^{11} + \overline{B}_S S+ \overline{B}^{I\!\!P} I\!\!P\right)
\end{equation}
in place of eq. (24). Eq. (A5) results in 
\begin{subequations}
\begin{align}
\frac{\partial \overline{B}_S}{\partial t} - \overline{A}_\delta &+ \overline{B}_S \left(\frac{g_{01}}{g_{11}}\right)_{,1}
- \overline{B}_{S,1} \left(\frac{g_{01}}{g_{11}}\right) \nonumber \\
 &+ \overline{B}_{I\!\!P} \left(\frac{-\sqrt{-g}}{g_{11}}\right)_{,1}  - \overline{B}_{I\!\!P,1} \left(\frac{-\sqrt{-g}}{g_{11}}\right) = 0\\
\intertext{\rm{and}}
 \nonumber \\
\frac{\partial \overline{B}_{I\!\!P}}{\partial t} - \overline{A}_\rho &+ \overline{B}_{I\!\!P} \left(\frac{g_{01}}{g_{11}}\right)_{,1}
- \overline{B}_{I\!\!P,1} \left(\frac{g_{01}}{g_{11}}\right)\nonumber\\
&+ \overline{B}_S \left(\frac{-\sqrt{-g}}{g_{11}}\right)_{,1} - \overline{B}_{S,1} \left(\frac{-\sqrt{-g}}{g_{11}}\right) = 0.
\end{align}
\end{subequations}
From eqs. (34) and (36) we see that $G_C = \overline{G}_{HTZ}$.

With the generator $G_{HTZ}$ of eq. (24), we find that 
\begin{align}
\delta f &= \left\lbrace f, G_{HTZ}\right\rbrace\nonumber \\
&= B_S p + B_{I\!\!P} f_{,1}
\end{align}
which by eq. (19) becomes 
\begin{equation}
= B_S \sqrt{-g}\, g^{00} f_{,0} + \left(B_S \sqrt{-g}\, g^{01} + B_{I\!\!P}\right)f_{,1}.
\end{equation}
This is identical to the diffeomorphism transformation
\begin{equation}
\delta f = \eta^0 f_{,0} + \eta^1f_{,1}
\end{equation}
provided
\begin{align}
B_S &= - \frac{\sqrt{-g}}{g_{11}} \eta^0\\
B_{I\!\!P} &= \eta^1 + \frac{g_{01}}{g_{11}} \eta^0.
\end{align}

Eq. (25) cannot be uniquely solved for $A_{11}$, $A_{00}$ and $A_{01}$ in terms of $B_S$ and $B_{I\!\!P}$, but a particular solution with $B_S$ and $B_{I\!\!P}$ given by eqs. (40, 41) is 
\begin{subequations}
\begin{align}
A_{11} &= 2g_{01} \eta_{,1}^0 + 2g_{11} \eta_{,1}^1 + \eta^0g_{11,0} + \eta^1 g_{11,1}\\
A_{00} &= 2g_{01} \eta_{,0}^1 + 2g_{00} \eta_{,0}^0 + \eta^1 g_{00,1} + \eta^0 g_{00,0}\\
A_{01} &= g_{00} \eta_{,1}^0 + g_{01} \left( \eta_{,0}^0 + \eta_{,1}^1\right) + g_{11}  \eta^1_{,0} \nonumber \\
&\quad+ \eta^0 g_{01,0} + \eta^1 g_{01,1}.
\end{align}
\end{subequations}
These expressions are consistent with $\delta g_{\mu\nu} = \left\lbrace g_{\mu\nu}, G_{HTZ}\right\rbrace$ giving the diffeomorphism transformation
\begin{equation}
\delta g_{\mu\nu} = g_{\mu\rho} \eta^\rho_{,\nu} + g_{\nu\rho} \eta_{,\mu}^\rho + \eta^\rho g_{\mu\nu ,\rho}.
\end{equation}
An additional solution to eq. (25) is
\begin{equation}
B_S = B_{I\!\!P} = 0
\end{equation}
\begin{equation}
A_{00} = \Lambda g_{00},\quad A_{11} = \Lambda g_{11}, \quad A_{01} = \Lambda g_{01}
\end{equation}
so that
\begin{equation}
\delta g_{\mu\nu} = \left\lbrace g_{\mu\nu} G_{HTZ}\right\rbrace = \Lambda g_{\mu\nu}.
\end{equation}
This is the Weyl conformal (scale) invariance. The transformations generated by $G_{HTZ}$ has also been found in ref. [23], and can also be 
found using $G_C$ and $\overline{G}_{HTZ}$.

We now consider gauge invariance in two dimensions when a massless scalar field is coupled to the metric and the EH action is first 
order.  Some aspects of this action were considered in ref. \cite{12}.

\section{First Order EH Action and Scalar Fields}

In $d$ dimensions, the action of eq. (8) can be written
\begin{equation}
S_{hG} = \int d^dx\, h^{\mu\nu} \left(G_{\mu\nu ,\lambda}^\lambda + \frac{1}{d-1} G_{\lambda\mu}^\lambda G_{\sigma\nu}^\sigma - G_{\sigma\mu}^\lambda G_{\lambda\nu}^\sigma\right).
\end{equation}
We begin by examining the equations of motion that follow from this form of the first order EH action before considering its canonical structure.
From eq. (47), the equations of motion for $G_{\mu\nu}^\lambda$ is 
\begin{align}
h^{\mu\nu}_{,\lambda} - \frac{1}{d-1} &\left(\delta_\lambda^\mu h^{\nu\alpha} + \delta_\lambda^\nu h^{\mu\alpha}\right) G_{\alpha\beta}^\beta\nonumber \\
&+ G_{\lambda\alpha}^\mu h^{\nu\alpha} + G_{\lambda\alpha}^\nu h^{\mu\alpha} = 0
\end{align}
from which it follows immediately that
\begin{equation}
G_{\alpha\beta}^\beta = -\frac{1}{2} \left(\frac{d-1}{d-2}\right)h_{\rho\sigma}h^{\rho\sigma}_{,\alpha}.
\end{equation}
Substitution of eq. (49) into eq. (48) gives 
\begin{align}
h_{,\lambda}^{\mu\nu} + \frac{1}{2(d-2)} &\left(\delta_\lambda^\mu h^{\nu\alpha} + \delta_\lambda^\nu h^{\mu\alpha}\right) h_{\rho\sigma}h^{\rho\sigma}_{\;,\alpha}\nonumber \\
&+ G_{\lambda\alpha}^\mu h^{\nu\alpha} + G_{\lambda\alpha}^\nu h^{\mu\alpha} = 0
\end{align}
which when combined with equations for $h^{\nu\lambda}_{\;,\mu}$ and $h^{\lambda\mu}_{\;,\nu}$ leads to
\begin{align}
G_{\mu\nu}^\lambda &= \frac{1}{2} h^{\lambda\rho}\left( h_{\mu\rho ,\nu} + h_{\nu\rho , \mu} - h_{\mu\nu , \rho}\right) \nonumber \\
&- \frac{1}{2(d-2)} h_{\mu\nu}h^{\lambda\rho}h_{\alpha\beta}h^{\alpha\beta}_{\;,\rho} .
\end{align}
For $d\neq2$, this is equivalent to having $\Gamma_{\mu\nu}^\lambda = \left\lbrace \begin{array}{c}
\lambda \\
\mu\nu \end{array}\right\rbrace$.~ 
From eqs. (49, 51) it is apparent that $d = 2$ dimensions is special.  If $d = 2$, then eq. (48) leads to a consistency condition on the equations of 
motion for $G_{\mu\nu}^\lambda$
\begin{equation}
h_{\mu\nu} h^{\mu\nu}_{\;,\lambda} = \frac{1}{\Delta} \Delta_{,\lambda} = 0 \quad (\Delta \equiv \det h^{\mu\nu})
\end{equation}
in place of eq. (49). Eq. (52) is consistent with 
\begin{equation}
\Delta = (\det h^{\mu\nu}) = - (-\det g_{\mu\nu})^{\frac{d}{2}-1}
\end{equation}
when $d = 2$.

If now we set
\begin{equation}
G_{\mu\nu}^\lambda  =  \frac{1}{2} h^{\lambda\rho}\left( h_{\mu\rho ,\nu} + h_{\nu\rho , \mu} - h_{\mu\nu , \rho}\right) + h_{\mu\nu} X^\lambda
\end{equation}
where $X^\lambda$ is an arbitrary vector, then 
\begin{align}
&- \left(\delta_\lambda^\mu h^{\nu\alpha} + \delta_\lambda^\nu h^{\mu\alpha}\right) G_{\alpha\beta}^\beta + G_{\lambda\alpha}^\mu h^{\nu\alpha} + 
G_{\lambda\alpha}^\nu h^{\mu\alpha}\nonumber \\
 &\qquad= - \frac{1}{2}\left(\delta_\lambda^\mu h^{\nu\alpha} + \delta_\lambda^\nu h^{\mu\alpha}\right) h^{\sigma\rho}h_{\sigma\rho ,\alpha} - h^{\mu\nu}_{\;,\lambda}
\end{align}
and hence eq. (54) satisfies eq. (48) provided eq. (52) is also satisfied.  Arbitrariness is also present in $\Gamma_{\mu\nu}^\lambda$ \cite{14,15} when $d = 2$ if the equation 
of motion for $\Gamma_{\mu\nu}^\lambda$ that follows from the first order form of the EH action in terms of $\Gamma_{\mu\nu}^\lambda$ and $g_{\mu\nu}$ is 
solved to give eq. (3).  Substitution of eq. (3) into the first order form of the EH action in terms of $\Gamma_{\mu\nu}^\lambda$ and $g_{\mu\nu}$ \
yields the second order form of the two dimensional EH action with all dependence on the arbitrary vector $\xi^\lambda$ dropping out.  In contrast, 
substitution of eq. (54) into eq. (47) with $d = 2$ leads to 
\begin{align}
\int dx^2 & [ h^{\mu\nu}  \left(G_{\mu\nu ,\lambda}^\lambda + G_{\lambda\mu}^\lambda G_{\sigma\nu}^\sigma - G_{\sigma\mu}^\lambda G_{\lambda\nu}^\sigma\right)] \\
&= \int dx^2\left[ \left(2X^\lambda + \frac{1}{2\Delta} h^{\lambda\rho} \Delta_{,\rho} + h^{\lambda\rho}h^{\sigma\tau}h_{\rho\sigma ,\tau}\right)_{,\lambda} \right. \nonumber \\
&\qquad - \frac{1}{\Delta} X^\lambda \Delta_{,\lambda} + \frac{1}{4\Delta^2} h^{\mu\nu} \Delta_{,\mu}\Delta_{,\nu}\nonumber \\
&\qquad   + \frac{1}{4} h^{\mu\nu} h_{,\mu}^{\alpha\beta} h_{\alpha\beta ,\nu} + \frac{1}{2} h_{\mu\nu} h^{\alpha\mu}_{\;\,,\beta} h^{\beta\nu}_{\;\;,\alpha}\Bigg].\nonumber
\end{align}
Upon dropping the total derivatives in eq. (56), we see that $X^\lambda$ remains as a Lagrange multiplier that ensures that eq. (52) is satisfied.
Thus the role of $X^\lambda$ in eq. (54) is different from that of $\xi^\lambda$ in eq. (3).

We now perform a canonical analysis of $S_{hG}$ when $d = 2$. In order to do this we rewrite eq. (47) as 
\begin{align}
S_{hr} &= \int d^2x \bigg[ -G_{00}^0 h_{,0} - 2G_{01}^0h_{,0}^1 - G_{11}^0h^{11}_{,0}\\
&\qquad\qquad - G_{00}^1 (h_{,1} + 2hG_{01}^0 + 2h^1G_{11}^0)\nonumber \\
&\qquad\qquad - 2G_{01}^1 (h_{,1}^1 - hG_{00}^0 + h^{11}G_{11}^0)\nonumber \\
 &\qquad\qquad - G_{11}^1 (h_{,1}^{11} - 2h^1G_{00}^0 - 2h^{11}G_{01}^0)\bigg] .\nonumber \\
&\qquad\qquad\qquad\qquad  (h = h^{00}, \quad h^1 = h^{01})\nonumber 
\end{align}
From eq. (57) it is apparent that the momenta conjugate to $(h, h^1, h^{11})$ are 
\begin{equation}
\pi = -G_{00}^0,\quad \pi_1 = -2G_{11}^0,\quad \pi_{11} = -G_{11}^0
\end{equation}
respectively.  The momenta conjugate to the ``Lagrange multiplier'' fields $(\xi^1 = G_{00}^1,\quad \xi = 2G_{01}^1,\quad
\xi_1 = G_{11}^1)$ are zero; these primary constraints lead to the secondary constraints
\begin{subequations}
\begin{eqnarray}
\phi_1 &=& h_{,1} - h\pi_1 - 2h^1\pi_{11}\\
\phi &=& h_{,1}^1 + h\pi - h^{11}\pi_{11}\\
\phi^1 &=& h_{,1}^{11} + 2h^1\pi + h^{11}\pi_{1}.
\end{eqnarray}
\end{subequations}
(These fields $\xi^1,\xi,\xi_1$ are in fact treated as degrees of freedom, and are not merely Lagrange multipliers as is done in refs. \cite{34,35}.)
This constraint structure leads to the gauge transformation of eqs. (4, 5) \cite{7,8,9,10,11,12}. We see that despite the fact that $G^1_{\mu\nu}$ is a ``Lagrange multiplier``
field, its transformation under eq. (5) is not merely an arbitrary shift, demonstrating why it needs to be treated as a dynamical variable whose associated canonical momentum vanishes. 
Under this transformation
\begin{equation}
\delta\Delta = 0
\end{equation}
and, according to eq. (54),
\begin{align}
\delta X^\mu &= \delta\left( h^{\mu\nu} G_{\lambda\nu}^\lambda - \frac{1}{2\Delta} h^{\mu\nu}\Delta_{,\nu}\right)\\
&= - h^{\mu\nu}\epsilon^{\lambda\sigma}\omega_{\nu\lambda ,\sigma} + \epsilon^{\mu\nu}\omega_{\nu\lambda} h^{\lambda\sigma}G_{p\sigma}^\rho\nonumber\\
& \qquad -h^{\mu\nu}G^\lambda_{\rho\nu}\epsilon^{\rho\sigma}\omega_{\lambda\sigma} - \frac{1}{2\Delta} \left(\epsilon^{\mu\lambda} h^{\sigma\nu} + 
\epsilon^{\nu\lambda}h^{\sigma\mu}\right)\Delta_{,\nu}.\nonumber
\end{align}

Let us now supplement the action of eq. (47) with $d = 2$ by
\begin{equation}
S_f = \frac{1}{2} \int dx^2\, h^{\mu\nu} f_{,\mu}f_{,\nu}.
\end{equation}
The canonical momenta if $h^{\mu\nu}$, $G_{\mu\nu}^\lambda$ and $f$ are all independent fields given by
\begin{equation}
p = \frac{\partial\mathcal{L}}{\partial f_{,0}} = hf_{,0} + h^1f_{,1}
\end{equation}
\begin{equation}
\Pi_\lambda^{\mu\nu} = \frac{\partial\mathcal{L}}{\partial G_{\mu\nu ,0}^{\lambda}} =0
\end{equation}
as well as ($\pi$, $\pi_1$ and $\pi_{11}$).

The canonical Hamiltonian is
\begin{equation}
\mathcal{H}_C = \frac{1}{h} \Sigma + \left(\frac{-h^1}{h}\right) I\!\!P + \xi^1\phi_1 + \xi\phi + \xi_1\phi^1,
\end{equation}
where 
\begin{equation}
\Sigma = \frac{1}{2}(p^2 - \Delta f_{,1}^2)
\end{equation}
and $I\!\!P$ is given in eq. (22). We now will show that $\phi^1,\,\phi,\,\phi_1,\,\\ I\!\!P$ and $\Sigma$ are all first class constraints.

The primary constraints
\begin{equation}
\Pi_1^{\mu\nu} = 0
\end{equation}
are first class; they lead to the secondary first class constraints
\begin{equation}
\phi_1 = \phi = \phi^1 = 0.
\end{equation}
One can show that
\begin{equation}
\left\lbrace \phi_1, \phi^1\right\rbrace = 2\phi\qquad , \left\lbrace \phi, \phi^1\right\rbrace = \phi^1, \qquad  \left\lbrace \phi_1, \phi\right\rbrace = \phi_1
\end{equation}
\begin{equation}
\left\lbrace \phi_1, \Delta\right\rbrace = \left\lbrace \phi, \Delta \right\rbrace =  \left\lbrace \phi^1, \Delta\right\rbrace = 0
\end{equation}
\begin{equation}
\Delta_{,1} = h \phi^1 + h^{11}\phi_1 - 2h^1 \phi,
\end{equation}
and, by using test functions as in ref. \cite{24},
\begin{equation}
\left\lbrace \Sigma(x), \Sigma(y)\right\rbrace = (\Delta(x) I\!\!P(x) \partial^y_1 - \Delta(y) I\!\!P(y)\partial_1^x)\delta(x-y)
\end{equation}
\begin{subequations}
This is not identical to the algebra of eq. (23a) unless $\Delta=1$. In addition we have
\begin{align}
&\left\lbrace \Sigma(x), I\!\!P(y)\right\rbrace  \nonumber \\
 &\quad=\left[(-\Sigma(x)\partial^y_1 + \Sigma(y)\partial_1^x) + \frac{1}{2} f_{,1}^2\Delta_{,1}\right]\delta(x-y)\\
&\left\lbrace I\!\!P(x), \Sigma(y)\right\rbrace  \nonumber \\
&\quad= \left[-\Sigma(x)\partial^y_1 + \Sigma(y)\partial_1^x - \frac{1}{2} f_{,1}^2\Delta_{,1}\right]\delta(x-y)
\end{align}
\end{subequations}
Only if $\Delta_{,1}=0$ does eq. (73) reduce to the algebra of eq. (23b)
for the tertiary first class constraints $\Sigma$ and $I\!\!P$.

As was the case when we considered coupling $N$ scalars to the metric field in section 2, the EH action by itself has no net physical degrees of freedom, 
while with the $N$ scalar fields there are $2N-4$ net physical degrees of freedom.

If the equation of motion were invoked so that by eq. (52) $\Delta$ would be constant, then $h$, $h^1$ and $h^{11}$ would not be independent, nor by eq. (71) would 
$\phi_1$, $\phi$ and $\phi^1$.  However, we will not impose this condition so that all components of $h^{\mu\nu}$ are independent.
(One could also ensure that $\Delta$ is constant by using a Lagrange multiplier.)

Using the HTZ approach, \cite{18,19} the generator of a gauge transformation is, by eq. (A2), of the form 
\begin{equation}
G = \int dx (a^1\Pi_1 + a\Pi +  a_1\Pi^1 + b^1\phi_1 + b\phi + b_1\phi^1 + c_\Sigma \Sigma + c_{I\!\!P}I\!\!P)
\end{equation}
where $\Pi_1$, and $\Pi$ and $\Pi^1$ are the momenta conjugate to $\xi^1$, $\xi$ and $\xi_1$ respectively.  By eqs. (65, 69-73) it follows that 
\begin{align}
&\left\lbrace G, \int dy \mathcal{H}_c \right\rbrace = \int dx \left\lbrace -a^1\phi_1 - a\phi - a_1\phi^1\right.\nonumber \\
&+ (b^1\xi - b\xi^1)\phi_1+2 (b^1\xi_1 - b_1\xi^1)\phi  + (b\xi_1 - b_1\xi)\phi^1\nonumber\\
&+ \frac{1}{h^2} (bh + 2b_1h^1)\Sigma\nonumber\\ 
&+ \frac{1}{h^2}\left[-hh^1b-h^2b^1 + (hh^{11} - 2h^{1{^2}})b_1\right]I\!\!P\nonumber \\
&+ \left[\Delta\left(c_{\Sigma,1}(\frac{1}{h}) - c_\Sigma(\frac{1}{h})_{,1}\right) + c_{I\!\!P,1}(\frac{h^1}{h}) - c_{I\!\!P,1}(\frac{h^1}{h})_{,1}\right]I\!\!P\nonumber \\
&+ \left[ c_{\Sigma,1}(\frac{h^1}{h}) - c_\Sigma(\frac{h^1}{h})_{,1} - c_{I\!\!P,1}(\frac{1}{h}) + c_{I\!\!P}(\frac{1}{h})_{,1}\right] \Sigma\nonumber \\
& \left. - \frac{1}{2} \Delta_{,1}f_{,1} \left(\frac{h^1}{h} c_\Sigma + \frac{1}{h} c_{I\!\!P}\right)\right\rbrace
\end{align}
provided we ignore possible dependence of $(a^1, a, a_1)$ and $(b^1, b, b_1)$ on dynamical variables.  (In the HTZ approach, ($c_\Sigma , c_{I\!\!P}$) are chosen to be 
independent of dynamical variables.)

Eq. (A5) to orders $\Sigma$ and $I\!\!P$ respectively gives 
\begin{equation}
\frac{\partial c_\Sigma}{\partial t} + \left[ + c_{\Sigma ,1} (\frac{h^1}{h}) - c_{\Sigma} (\frac{h^1}{h})_{,1} - c_{I\!\!P ,1} (\frac{1}{h})  + c_{I\!\!P} (\frac{1}{h})_{,1} \right]
\end{equation}
\begin{equation}
+ \frac{1}{h^2} (bh + 2b_1h^1) = 0\nonumber
\end{equation}
\begin{equation}
\frac{\partial c_{I\!\!P}}{\partial t} + \left[ \Delta\left( c_{\Sigma ,1} (\frac{1}{h}) - c_{\Sigma} (\frac{1}{h})_{,1}\right) + c_{I\!\!P ,1} (\frac{h^1}{h})  - c_{I\!\!P}
 (\frac{h^1}{h})_{,1}\right]
\end{equation}
\begin{equation}
+ \frac{1}{h^2} [-hh^1b -h^2b^1 + (hh^{11} - 2h^{1^{2}})b_1] = 0\nonumber
\end{equation}
which relate $(b^1, b, b_1)$  to ($c_\Sigma , c_{I\!\!P}$).  These equations are altered when ($c_\Sigma , c_{I\!\!P}$) depend on $(h, h^1, h^{11})$ by terms linear in 
$(\xi^1, \xi , \xi_1)$.

We find that much like eq. (38)
\begin{equation}
\delta f = \left\lbrace f,G\right\rbrace = (c_\Sigma h) f_{,0} + (c_\Sigma h^1 + c_{I\!\!P})f_{,1}
\end{equation}
which reduce to eq. (39) provided $c_\Sigma$ and $c_{I\!\!P}$ acquire dependence on $h^1$ and $h$,
\begin{equation}
c_\Sigma = \eta^0/h
\end{equation}
\begin{equation}
c_{I\!\!P} = \eta^1 - h^1 \eta^0/h.
\end{equation}
If $c_\Sigma$ and $c_{I\!\!P}$ have this form, then eqs. (76) and (77) acquire extra contributions on the left side of 
\begin{equation}
- \frac{\eta^0\xi}{h} - \frac{2h^1\eta^0\xi_1}{h^2}
\end{equation}
and 
\begin{equation}
\frac{h^1\eta^0}{h} \xi + \eta^0\xi^1 + \frac{1}{h^2} (2h^{1^{2}} - hh^1)\eta^0\xi_1
\end{equation}
respectively.  Upon substituting eqs. (79, 80) into eqs. (76, 77) when supplemented by eqs. (81, 82) we find two equations for $b$, $b_1$ and $b^1$ that are consistent 
with taking 
\begin{align}
b &= \eta_{,0}^0 + \eta_{,1}^1 + \eta^0\xi\\
b_1 &= \frac{1}{2h^1}\left(\eta^0 h_{,0} + \eta^1 h_{,1} - 2h^1 \eta_{,1}^0 - 2h\eta_{,0}^0 \right) + \eta^0\xi_1\\
b^1 &= \frac{1}{h^1}\left(\eta^1_{,0} h^1 - \eta_{,0}^0 h^{11}\right)+ \frac{h^{11}}{2hh^1} \left(\eta^1h_{,1} + \eta^0h_{,0}\right)\\
 & \qquad \qquad - \frac{1}{h} (\eta^1 h_{,1}^1 + \eta^0 h_{,0}^1) + \eta^0\xi^1.\nonumber
\end{align}
With ($b, b_1, b^1$) given by eqs. (83-85) we find that 
\begin{align}
\delta h &= \left\{h,G\right\} = -h \eta_{,0}^0 + h\eta_{,1}^1 + \eta^0 h_{,0} + \eta^1h_{,1} \\
 & \qquad\qquad\qquad - 2h^1 \eta_{,1}^0+\eta^0 (h\xi + 2h^1\xi_1)\nonumber\\
\delta h^1 &= \left\{h^1,G\right\} = -h \eta_{,0}^1 + \eta^1h_{,1}^1  + \eta^0 h_{,0}^1 - h^{11}\eta_{,1}^0\\
&  \qquad\qquad\qquad\qquad  + \eta^0 (-h\xi^1 + h^{11}\xi_1)\nonumber\\
\delta h^{11} &= \left\{h^{11},G\right\} = -2h^1 \eta_{,0}^1 + h^{11} \eta_{,0}^0 -  h^{11}\eta_{,1}^1  \\
 & + h^{11}_{,0}\eta^0 + h^{11}_{,1}\eta^1-\frac{1}{h}(\Delta_{,0} \eta^0 + \Delta_{,1}\eta^1) \nonumber\\
&\qquad\qquad\qquad + \eta^0 (-2h^1\xi^1 - h^{11}\xi). \nonumber
\end{align}

From eq. (43), under a diffeomorphism transformation
\begin{equation}
\delta h^{\mu\nu} = h^{\mu\lambda} \theta_{,\lambda}^\nu + h^{\nu\lambda} \theta_{,\lambda}^\mu - (h^{\mu\nu}\theta^\lambda)_{,\lambda}
\end{equation}
which is the transformation of eqs. (86-88) provided
\begin{equation}
\theta^\lambda = -\eta^\lambda \;\;,\qquad \Delta_{,0} = \Delta_{,1} = 0\;\; {\rm{and}}\;\;\xi^1 = \xi = \xi_1 = 0.\nonumber
\end{equation}

An additional solution to eqs. (76, 77) is 
\begin{equation}
c_\Sigma = c_{I\!\!P} = 0 \;,\qquad b = \frac{-2b_1h^1}{h}\;, b^1 = \frac{h^{11}b_1}{h}
\end{equation}
so that
\begin{equation}
b^1\phi_1 + b\phi + b_1\phi^1 = \frac{b_1}{h} \Delta_{,1}\;,
\end{equation}
and hence
\begin{equation}
\delta h^{\mu\nu} = \left\{ h^{\mu\nu},G\right\} = 0.
\end{equation}

Finding the variation of $G_{\mu\nu}^\lambda$ requires knowing the coefficients ($a^1, a, a_1$) in eq. (74).  These are found by considering 
these terms in eq. (A5) proportional to ($\phi^1, \phi, \phi_1$).  By eq. (75), these are respectively given by 
\begin{subequations}
\begin{align}
\frac{\partial b_1}{\partial t} &- a_1 + (b\xi_1 - b_1\xi) - \frac{1}{2} f_{,1}^2 (h^1c_\Sigma + c_{I\!\!P}) = 0\\
\frac{\partial b}{\partial t} &- a + 2(b^1\xi_1 - b_1\xi^1) +  f_{,1}^2 \frac{h^1}{h}(h^1c_\Sigma + c_{I\!\!P}) = 0\\
 \frac{\partial b^1}{\partial t}&- a^1 +(b^1\xi - b\xi^1) \\
&-\frac{1}{2} f_{,1}^2 \frac{h^{11}}{h}(h^1c_\Sigma + c_{I\!\!P}) = 0  \nonumber
\end{align}
\end{subequations}
provided we ignore terms in $\left\{G, \mathcal{H}_c\right\}$ that are linear in ($\phi^1, \phi , \phi_1$) on account of the dependency of 
($b^1, b, b_1$) on ($h, h^1, h^{11}$) following from eqs. (76, 77).  If one were to supplement eqs. (92, 93) with terms 
\begin{equation}
\phi^1 \left\{ b_{1,}\phi^1\xi_1 + \phi\xi + \phi_1\xi^1\right\} + \phi\left\{ b, \phi^1\xi_1 + \phi\xi + \phi_1\xi^1\right\}\nonumber
\end{equation}
\begin{equation}
+ \phi_1 \left\{ b^1, \phi^1\xi_1 + \phi\xi + \phi_1\xi^1\right\}
\end{equation}
in order to take into account the dependency of ($b_1, b, b^1$) on ($h, h^1, h^{11}$), and use eqs. (83-85) for ($b_1, b, b^1$), one 
encounters ill defined PBs of the form $\left\{h_{,0}, \pi\right\}$ indicating a breakdown of the HTZ procedure for finding the generator of a gauge transformation 
that leads to eq. (A5).

However, it is possible to overcome this shortcoming of the HTZ approach for finding the generator of a gauge transformation.  If instead of eqs. (A3), one were 
to take the change in a dynamical variable $A$ to be given by 
\begin{equation}
\delta A = \nu^{a_{i}} \left\{ A, \gamma_{a_{i}}\right\}
\end{equation}
so that $\nu^{a_{i}}$ is not affected when one computes the PB,
then the change in the extended action of eq. (A1) would be 
\begin{align}
\delta S_E = & \int dt \bigg[- v^{a{_{i}}} \big(\left\{\gamma_{a_{i}},p^j\right\}\dot{q}_j - \left\{\gamma_{a_{i}},q_j\right\}\dot{p}^i\\
&- \left\{ \gamma_{a_{i}},q_j\right\} \frac{\partial H_c}{\partial q_i} - \left\{ \gamma_{a_{i}},p^j\right\} \frac{\partial H_c}{\partial p^j} \nonumber\\
&  - U^{a_{j}} \left\{ \gamma_{a_{i}},\gamma_{a_{j}}\right\} \big) - \delta U^{a_{i}} \gamma_{a_{i}}\bigg]\nonumber
\end{align}
provided we do an integration by parts, dropping the surface term.  Eq. (96) further reduces to 
\begin{align}
\delta S_E =  \int dt & \bigg[- v^{a{_{i}}} \big(\frac{\partial\gamma_{a_{i}}}{\partial q_j}\dot{q}_j + \frac{\partial\gamma_{a_{i}}}{\partial p^j}\dot{p}^j\\
&- \left\{ \gamma_{a_{i}},H_c + U^{a_{j}} \gamma_{a_{j}}\right\} \big) - \delta U^{a_{i}}\gamma_{a_{j}}\bigg]\nonumber
\end{align}
as $u^{a_{j}}$ is not dynamical; a further integration by parts without keeping the surface terms leads to 
\begin{equation}
\delta S_E = \int dt \Bigg[ + \gamma_{a_{i}} \frac{D\nu^{a_{i}}}{Dt} + \nu^{a_{i}} \left\{\gamma_{a_{i}}, H_c + U^{a_{j}}\gamma_{a_{j}}\right\}  - \delta U^{a_{i}}\gamma_{a_{i}}\Bigg]
\end{equation}
which is almost identical to eq. (A4).  However, the coefficients $\nu^{a_{i}}$ are not involved in the evaluation of any PBs.

For the system we have been considering, we can employ eq. (98) to find the gauge transformation of a dynamical variable $A$
\begin{align}
\delta A &= \overline{a}^1 \left\{ A,\Pi_1\right\} + \overline{a} \left\{ A,\Pi\right\} + \overline{a}_1 \left\{ A,\Pi^1\right\}\\
&\qquad\qquad + \overline{b}^1 \left\{ A,\phi_1\right\} + \overline{b} \left\{ A,\phi\right\} + \overline{b}_1 \left\{ A,\phi^1\right\}\nonumber \\
& \qquad\qquad\qquad + \overline{c}_\Sigma \left\{ A,\Sigma \right\}+ \overline{c}_{I\!\!P} \left\{ A,I\!\!P\right\}.\nonumber
\end{align}
Eq. (98), when used in the same way eq. (A4) has been used by HTZ \cite{18,19} fixes ($\overline{b}^1, \overline{b}, \overline{b}_1$) in terms of 
($\overline{c}_{\Sigma}, \overline{c}_{I\!\!P}$) by eqs. (76, 77) and in turn determines ($\overline{a}^1, \overline{a}, \overline{a}_1$) by 
eqs. (92, 93).

We find that, for example, that eq. (95) leads to 
\begin{equation}
\hspace{-6cm}\delta G_{01}^1 = \overline{a}\left\{\frac{1}{2} \xi, \Pi\right\}
\end{equation}
which, by eq. (93b) becomes 
\begin{equation}
\qquad = \frac{1}{2} \left[ \frac{\partial\overline{b}}{\partial t} + 2 (\overline{b}^1 \xi_1 - \overline{b}_1 \xi^1) + f_{,1}^2 \frac{h^1}{h}(h^1 \overline{c}_\Sigma + \overline{c}_{I\!\!P})\right].
\end{equation}
Eqs. (79, 80, 83-85) in turn show that eq. (101) reduces to 
\begin{subequations}
  \begin{eqnarray}
\delta G_{01}^1 &=& \frac{1}{2} \left[ \left(\eta_{,0}^0 + \eta_{,0}^1 + 2\eta^0 G_{01}^1\right)_{,0} + 2( \frac{1}{h^1} (\eta_{,0}^1 h^1 - \eta_{,0}^0h^{11})\right.\nonumber\\
&+& \frac{h^{11}}{2hh^1}(\eta^1 h_{,1}+ \eta^0 h_{,0}) - \frac{1}{h} (\eta^1 h^1_{,1} + \eta^0 h^1_{,0}))G_{11}^1\nonumber\\
&-& \left(\frac{1}{h^1}\right)  \left(\eta^0 h_{,0}+ \eta^1 h_{,1} - 2h\eta^0_{,1} - 2h\eta_{,0}^0\right)G_{00}^1\nonumber\\
 &+& \left. f_{,1}^2 \frac{h^1}{h}\eta^1\right]
   \end{eqnarray}
Similarly, we find that
   \begin{eqnarray} 
&\delta G_{00}^1& = \overline{a}_1\left\{\xi^1,\Pi_1\right\}\nonumber\\
&=& \frac{\partial\overline{b}_1}{\partial t} + (\overline{b}\xi_1 - \overline{b}_1 \xi) - \frac{1}{2} f_{,1}^2 (h^1c_\Sigma + c_{I\!\!P})
   \end{eqnarray}
and 
   \begin{eqnarray}
\delta G_{11}^1&=& \overline{a}^1\left\{\xi_1,\Pi^1\right\}= \frac{\partial\overline{b}^1}{\partial t} + (\overline{b}^1\xi - \overline{b} \xi^1)\nonumber \\
&-& \frac{1}{2} f_{,1}^2 \frac{h^{11}}{h}(h^1c_\Sigma + c_{I\!\!P})
   \end{eqnarray}
\end{subequations}
Eqs. (102) have a term proportional to $f_{,1}^2$; similarly by eqs. (95, 66), $\delta G_{00}^0$ has a term proportional to $-\frac{1}{2} h^{11} c_\Sigma f_{,1}^2$.  It is 
apparent that $\delta G_{\mu\nu}^\lambda$ always has a contribution proportional to $f_{,1}^2$.  This mixing of the affine connection and scalar field under a gauge transformation 
is somewhat unusual.  The change in $G_{\mu\nu}^\lambda$ under a diffeomorphism is 
\begin{align}
\delta G_{\mu\nu}^\lambda &= - G_{,\mu\nu}^\lambda + \frac{1}{2}\left(\delta_\mu^\lambda \theta_{,\nu\rho}^\rho + \delta_\nu^\lambda \theta_{,\mu\rho}^\rho\right) 
- \theta^\rho G_{\mu\nu ,\rho}^\lambda\nonumber \\
 &+ G_{\mu\nu}^\rho \theta_{,\rho}^\lambda - \left( G_{\mu\rho}^\lambda \theta_{,\nu}^\rho + 
G_{\nu\rho}^\lambda  \theta_{,\mu}^\rho\right)
\end{align}
which does not mix $G_{\mu\nu}^\lambda$ and $f_{,1}$.

It is also possible to use the approach of [16,17] to find the gauge generator associated with $S_{hG} + S_f$ when $d = 2$.  In eq. (A12), $N = 2$ since there are 
tertiary constraints.  With $G_2 = \Pi^1$ and $\mathcal{H}_c$ given by eqs. (65), it follows from 
\begin{equation}
G_1 + \left\{ G_2, H_c\right\} \approx p.c.
\end{equation}
that 
\begin{align}
G_1(x) &= \phi^1(x) + \int dy [\alpha^1 (x-y)\Pi_1(y)\nonumber \\
&+ \alpha(x-y)\Pi(y)+\alpha_1(x-y)\Pi^1(y)];
\end{align}
next
\begin{equation}
G_0 + \left\{ G_1, H_c\right\} \approx p.c.
\end{equation}
leads to 
\begin{align}
G_0 &= \int dy \left[\beta^1(x-y)\Pi_1(y) \right.\nonumber\\
& + \beta(x-y)\Pi(y)  +\beta_1(x-y)\Pi^1(y)\nonumber \\
&  \left. + \alpha^1(x-y)\phi_1(y) + \alpha(x-y)\phi(y) + \alpha_1(x-y)\phi^1(y)\right]\nonumber \\
&+ 2\xi^1(x)\phi(x) + \xi(x)\phi^1(x)\\
& \quad - \frac{2h^1(x)\Sigma(x)}{h^2(x)} + \left( \frac{2h^{1^{2}}(x) - h(x)h^{11}(x)}{h^2(x)}\right)I\!\!P(x).\nonumber
\end{align}
The final condition
\begin{equation}
 \left\{ G_0, H_c\right\} \approx p.c.
\end{equation}
is satisfied to orders $\Sigma$, $I\!\!P$, $\phi^1$, $\phi$ and $\phi_1$ respectively provided
\begin{subequations}
  \begin{align}
&\frac{\alpha}{h} + \frac{2h^1\alpha_1}{h^2} + \frac{4\xi^1}{h} + \frac{6 h^1 \xi}{h^2} + \left(\frac{8h^{1^{2}}- 2hh^{11}}{h^3}\right)\xi_1\nonumber \\
&- 2\left(\frac{h^{1^{2}}}{h^2}\right)_{,1} \frac{1}{h} + \left(\frac{hh^{11}}{h}\right)_{,1} \frac{1}{h^2} = 0
  \end{align}
  \begin{align}
 &-\alpha^1 - \frac{h^1\alpha}{h} + \left(\frac{-2h^{1^{2}}+ hh^{11}}{h^2}\right)\alpha_1 - \frac{4h^1\xi^1}{h}\nonumber \\
 &+ \left(\frac{-6h^{1^{2}}+ 3hh^{11}}{h^2}\right)\xi+ \left(\frac{-8h^{1^{2}}+ 6hh^{11}}{h^3}\right)(h^1\xi_1) \nonumber \\
 &- \frac{h^{11}}{h} \left(\frac{h^1}{h}\right)_{,1} + \left(\frac{2h^{1^{2}}- hh^{11}}{h^2}\right)_{,1} \left(\frac{h^1}{h}\right) = 0 
 \end{align}
 \end{subequations}
\begin{subequations}
  \begin{align}
&-\beta_1 + \alpha\xi_1 - \alpha_1\xi + 2\xi^1\xi_1 - \xi^2 + \frac{h^{11}}{2h} f_{,1}^2 \\
&\qquad\qquad\qquad+ \left\{\alpha_{,1}H_c\right\}  = 0\nonumber\\
&-\beta + 2(\alpha^1\xi_1 - \alpha_1\xi^1 -\xi\xi^1) - \frac{h^1h^{11}}{h^2} f_{,1}^2 \\
&\qquad\qquad\qquad+ \left\{\alpha,H_c\right\}  = 0\nonumber\\
&-\beta^1 + \alpha^1\xi - \alpha\xi^1 -2\xi^{1^{2}} + \frac{h^{11^{2}}f_{,1}^3}{2h^2}\\
&\qquad\qquad\qquad+ \left\{\alpha^1,H_c\right\} = 0.\nonumber
\end{align}
\end{subequations}
In exactly, the same way we find that if $G_2 = \Pi$, then
\begin{align}
G_1 &= \phi + \int dy (\alpha^1 \Pi_1 + \alpha\Pi + \alpha_1\Pi^1)\\
G_0 &= \int dy \left[\beta^1 \Pi_1 + \beta\Pi + \beta_1\Pi^1 + \alpha^1\phi_1 + \alpha\phi + \alpha_1\phi^1\right]\nonumber\\
&\qquad\qquad + \xi^1\phi_1 - \xi_1\phi^1 - \frac{1}{h} (\Sigma - h^1I\!\!P)
\end{align}
with 
\begin{subequations}
 \begin{align}
&\frac{\alpha}{h} + \frac{2h^1\alpha_1}{h^2} + \frac{\xi}{h} = 0\\
 &-\frac{\alpha^1}{h} - \frac{h^1\alpha}{h} + \frac{-2h^{1^{2}} + hh^{11}}{h^2} \alpha_1\\
&\qquad\qquad\qquad - 2\xi^1 - \frac{h^1\xi}{h} = 0\nonumber
\end{align}
\end{subequations}
\begin{subequations}
 \begin{align}
 &- \beta_1 + \xi_1\alpha - \xi\alpha_1 + \xi\xi_1 + \left\{\alpha_1,H_c\right\} = 0\\
  &- \beta + 2\left( \xi_1\alpha^1 - \xi^1\alpha_1 + 2\xi^1\xi_1\right) + \left\{\alpha,H_c\right\} = 0\\
 &- \beta^1 + \left( \xi\alpha^1 - \xi^1\alpha + \xi\xi^1\right) + \left\{\alpha^1,H_c\right\} = 0.
 \end{align}
 \end{subequations}
 Finally, if $G_2 = \Pi_1$, then we find that 
\begin{equation}
G_1 = \phi_1 + \int dy\left[ \alpha^1 \Pi_1 + \alpha\Pi + \alpha_1\Pi^1\right]
\end{equation}  
\begin{align}
G_0 &= \int dy \left[ \beta^1\Pi_1 + \beta\Pi + \beta_1\Pi^1 + \alpha^1\phi_1 +  \alpha\phi + \alpha_1\phi^1\right]\nonumber\\
&\qquad\qquad - \xi\phi_1 - 2\xi_1\phi + I\!\!P
\end{align}
and so 
\begin{subequations}
\begin{align}
 &\frac{\alpha}{h}+ \frac{2h^1}{h^2}\alpha_1 - \frac{2\xi_1}{h} + \left(\frac{1}{h}\right)_{,1} = 0\\
&- \alpha^1 + \alpha_1 \left( \frac{-2h^{1^{2}} + hh^{11}}{h^2}\right) - \frac{h^1}{h} \alpha+ \xi \nonumber \\
&+ \frac{2h^1}{h}\xi_1 - \left(\frac{h^1}{h}\right)_{,1} = 0
\end{align}
\end{subequations}
\begin{subequations}
 \begin{align}
 &- \beta_1 + \xi_1\alpha - \xi\alpha_1 - 2\xi_1^2 - \frac{1}{2} f_{,1}^2
 + \left\{\alpha_1,H_c\right\} = 0\\
  &- \beta +  2\xi_1\alpha^1 - 2\xi^1\alpha_1 - 2\xi\xi_1 + \frac{h^1}{h} f_{,1}^2\nonumber \\
 &\qquad\qquad\qquad\qquad+\left\{\alpha,H_c\right\} = 0\\
 &- \beta^1 + \xi\alpha^1 - \xi^1\alpha + 2\xi^1\xi_1 - \xi^2 - \frac{h^{11}}{2h} f_{,1}^2\nonumber \\ 
 &\qquad\qquad\qquad\qquad+\left\{\alpha^1,H_c\right\} = 0.
 \end{align}
 \end{subequations}
In the instance where $G_2 = \Pi^1$, the two conditions of eqs. (109a,b) do not fix $\alpha^1$, $\alpha$ and $\alpha_1$ uniquely; however eqs. 
(110a,b,c) do determine $\beta^1$, $\beta$ and $\beta_1$ in terms of $\alpha^1$, $\alpha$ and $\alpha$.  This lack of uniqueness in the gauge 
generator is a consequence of there being but two tertiary first class constraints following from the three primary first class constraints.  The 
same pattern is repeated when $G_2 = \Pi$ (eqs. (113, 114)) and $G_2 = \Pi$, (eqs. (117, 118)).  In each case though, $\beta^1$, $\beta$ and $\beta_1$ 
depend on $f_{,1}^2$ in such a way that the transformation $\delta G_{\mu\nu}^\lambda$ depends on $f_{,1}^2$ as was the case when the HTZ approach to 
finding a gauge generator was used.

\section{Discussion}

In this paper we have closely followed the Dirac constraint formalism \cite{1,2,3,4,5,6} to analyze the gauge structure of a two dimensional massless scalar 
field in curved space.  Though it has long been recognized that this is related to the Bosonic string \cite{21} and that this is a system involving constraints, it does not appear that 
a full constraint analysis has been performed on this system.  It always appears that some fields have been eliminated by choosing to work in a ``convenient'' gauge before 
the constraints are identified, or that the generator of gauge transformations is postulated rather than derived from the first class constraints (see for example 
ref. \cite{25}).

In this analysis we have included the EH action in second order form \cite{7}, even though it normally is dropped since it does not contain any dynamical degrees of freedom.  
This suggests that we also consider the first order EH action whose canonical structure in the absence of matter leads to a gauge invariance generated 
by the first class constraints that appears distinct from diffeomorphism invariance, and which accounts for the absence of dynamical degrees of freedom \cite{8,9,10,11,12,13}.  (We 
might also look at other actions for the two dimensional metric field be considered, such as the Weyl scalar invariant action which involves a vector field \cite{27}.)  
One peculiarity in our canonical analysis is that by adding the scalar field $f$, two degrees of freedom are added in phase space, but this also results in two more first class 
constraints (either $S$ and $I\!\!P$ or $\Sigma$ and $I\!\!P$ for the second order and first order EH actions respectively) which when combined with the 
associated gauge conditions, leads to a negative number of degrees of freedom ($-2$) in phase space. This issue was raised but not satisfactorily resolved in 
ref. \cite{12}. If there are $N$ scalars $f^a$ and the kinetic term for these scalars were $O(N)$ symmetric, then there are ten restrictions on $2N+6$ fields in phase space, 
leaving $2N-4$ independent degrees of freedom. There are also $2N-4$ net degrees of freedom when using the first order form of the EH action.

The problem with having an unexpected number of degrees of freedom (especially when $N=1$) is implicit in all discussions of the canonical structure of the Bosonic string
that we have encountered in the literature (see for example ref. \cite{25}) but no satisfactory resolution of the problem has been provided. In particular, if $N=1$, 
it would seem that the first class constraints of eqs. (21,22), or eqs. (22,66) would require imposing a gauge fixing that would over determine $f$ and its conjugate momentum $p$. 
For $N=26$ there is a positive number of degrees of freedom (48) even after a gauge is chosen and this problem of over determination of $f^{(a)}$ and $p^{(a)},(a=1...26)$ does not arise. 
Consequently, the Bosonic string does not suffer from this particular inconsistency. In fact though, one should not be surprised that if $N=1$ there are no degrees of freedom associated with the scalar $f$,
as the equation of motion for $h^{\mu\nu}$ that follows from eq. (7) is $(\partial_{\mu}f)(\partial_{\nu}f)=0$ which implies that $f$ does not propagate. The equation of motion that follows from $g_{\mu\nu}$
in eq. (2) is $\partial_{\mu}\partial_{\nu}f-\frac{1}{2}g_{\mu\nu}g^{\alpha\beta}\partial_{\alpha}f\partial_{\beta}f=0$ which has the same implications. For $N>1$fields, $f^{(a)}$ is not necessarily a constant 
in order to satisfy the equations of motion for the metric. 

Our analysis displays some interesting features of the approaches of C and HTZ to finding the gauge generator from the first class constraints.  First of all, it 
is apparent from our discussion of the gauge generator when the EH action is second order that the actual form of the generator is dependent on how the constraints 
are chosen.  When using the method of C, which form of the primary constraints is chosen is important (as was pointed out in ref. \cite{26}) while the form of the gauge 
generator found using the approach of HTZ is different when different linear combinations of constraints of the highest order are employed.

The diffeomorphism invariance manifestly present in the initial Lagrangian is only recovered when using the second order form of the EH action if the gauge parameters 
associated with the secondary constraints are field dependent (which is contrary to the HTZ approach).  There is also a residual symmetry occurring in this case. This 
additional symmetry resulting from the gauge generator is the Weyl scale symmetry. Thus both diffeomorphism invariance and Weyl scale invariance are gauge symmetries.
 
The HTZ formalism, when applied to first order form of the EH action plus the action for a scalar field, yields the diffeomorphism transformation for the scalar field only if the gauge 
parameters associated with the tertiary constraints are again field dependent.  The resulting equations for the gauge parameters associated with primary constraints 
involves ill defined PBs that can be avoided by slightly modifying the HTZ procedure.  When this is done, the resulting gauge transformation is unusual as it mixes the affine 
connection and the scalar field in an non-polynomial fashion.  We have attempted unsuccessfully to find such a gauge invariance directly from the action 
given in eqs. (47, 62).

Of course, once the canonical structure of these models is disentangled, their quantization is to be considered.  This may have implications for Bosonic string theory.

\begin{acknowledgement}
We would like to thank S.V. Kuzmin and N. Kiriushcheva for helpful discussions and R. Macleod for encouragement.
\end{acknowledgement}

\noindent
{\Large\bfseries{Appendix. The Gauge Generator}}

 When one is presented with a Lagrangian $L(q_i(t), \dot{q}_i(t))$, passing to the Hamiltonian formalism is straightforward unless the equations defining the 
canonical momenta $p^i = \partial L(q_i,\dot{q}_i)/\partial\dot{q}_i$ cannot be solved for $\dot{q}_i$ in terms of $q_i$ and $p^i$.  In this case, one must use 
the Dirac constraint formalism \cite{1,2,3,4,5,6}.  (For a discussion of the history of the constraint formalism, see ref. \cite{28}.)

If one encounters first class constraints $\gamma_{a_{j}}$ in the $j^{th}$ generation\footnote[2]{We assume all second class constraints have been 
used to eliminte some of the degrees of freedom and that the Dirac Brackets (DB) for the remaining variables are identical to their Poisson Brackets (PB).}, then the 
``extended action'' is 

\[ S_E  = \int_{t_i}^{t_f} dt\left[ p^i\dot{q}_i - H_c(q_i,p^i) - U^{a_{j}} (t)\gamma_{a_{j}}(q_i,p^i)\right];\eqno(A1)\]
where $H_c$ is the canonical Hamiltonian. (Primary constraints are of the first "generation", secondary constraints are of the second "generation" etc.) If a gauge generator $G$
is a linear combination of first class constraints as in the HTZ approach \cite{18,19}
\[ G  = \mu^{a_{j}}\left(q_i(t),p^i(t), U^{\alpha_{j}}(t),t\right)\gamma_{a_{j}}(q_i(t),p^i(t))\eqno(A2)\]
so that the change in a dynamical variable $A$ is given by the PB
\[\delta A = \left\{A,G\right\} \eqno(A3)\]
then this results in
\[\delta S_E  = \int_{t_i}^{t_f} dt\left[\frac{D\mu^{a_{j}}}{Dt} \gamma_{a_{j}} + \left\{G, H_c + U^{a_{j}}\mu_{a_{j}}\right\} - \delta 
U^{a_{j}}\gamma_{a_{j}}\right]\eqno(A4)\]
where $\delta U^{a_{j}}$ is the corresponding change in the Lagrange multiplier $U^{a_{j}}$ and $D/Dt$ is the time derivative induced by the implicit time 
dependence through $U^{a_{j}}(t)$ and the explicit time dependence. (The time dependence of $\gamma^{a_{j}}$  through $p^i(t)$ and $q^i(t)$ is canceled by the PB
$\lbrace \int dt p^i \dot{q_i},\mu^{a_j}\rbrace$.) 
Surface terms at $t=t_i,t_f$ are dropped  in eq. (A4).

One can move from the extended action of eq. (A1) to the ``total action'' $S_T$ by setting $U^{a_{j}} = \delta U^{a_{j}} = 0$ for $j \geq 2$.  This total action has the 
same invariance as the classical action $\int dt L$ \cite{29}.  Consequently one can find invariance of the classical action by determining the functions $\mu^{a_{j}} (j = 1, 2 \ldots N)$ 
in eq. (A2) by solving
\[\frac{D\mu^{a_{j}}}{Dt} \gamma_{a_{j}} + \left\{G, H_c + U^{a_{1}}\gamma_{a_{1}}\right\} - \delta U^{a_{1}}\gamma_{a_{1}} = 0\eqno(A5)\]
systematically; as eq. A(5) ensures that $S_T$ remains invariant; $\mu^{a_{N}}$ is taken to be an arbitrary function of time, $\mu^{a_{N-1}}$ is fixed in terms of $\mu^{a_{N}}$; $\mu^{a_{N-2}}$ 
is fixed in terms of
$\mu^{a_{N-1}}$ etc.

An approach to finding the gauge invariance in a system with a denumerable number of degrees of freedom in which the Lagrangian is at most linear in time derivatives appears in refs. \cite{34,35}. 
However, this discussion does not consider the possibility of tertiary constraints (which occur in the first order form of the EH action in $D>2$ dimensions \cite{36,37}) nor is it extendable 
to deal with such constraints. Also, it does not exploit the fact that it is the total action,
not the extended action, that has the same invariances as the classical action in order to find dependence of the gauge transformation on the time derivative of the gauge functions.

In the approach of C \cite{16,17}, the generator $G$ is found by considering the Hamiltonian equations of motion.  If both $(q_i, p^i)$ and $(q_1 + \alpha_i, p^i + \beta^i)$ 
are solutions, then 
\[\alpha_i = \left\{q_i, G\right\} = \frac{\partial G}{\partial p^i},\qquad \beta^i = \left\{p^i, G\right\} = - \frac{\partial G}{\partial q_i}.\eqno(A6) \]
We now have the general equation $\frac{dA}{dt} \approx \left\{A, H_T\right\} + \frac{\partial A}{\partial t}$ which means that eq. (A6) leads to 
\[\dot{\alpha}_i \approx \left\{ \frac{\partial G}{\partial p^i}, H_T \right\} + \frac{\partial^2 G}{\partial t\partial p^i}, \quad 
\dot{\beta}^i \approx -\left\{ \frac{\partial G}{\partial q_i}, H_T \right\} - \frac{\partial^2 G}{\partial t\partial q_i}.\eqno(A7)\]
(The weak inequality $A \approx B$ holds when the primary constraints vanish.)  In addition, the equations of motion themselves lead to 
\[ \dot{q}_i + \dot{\alpha}_i \approx \frac{\partial}{\partial p^i} H_T (q_i + \alpha_i, p^i + \beta^i),\]
\[ \dot{p}^i + \dot{\beta}^i \approx - \frac{\partial}{\partial q_i}
H_T (q_i + \alpha_i, p^i + \beta^i)\eqno(A8)\]
which to lowest order becomes 
\[ \dot{\alpha}_i \approx \frac{\partial}{\partial p^i} \left( \frac{\partial H_T}{\partial q_i} \alpha_i + \frac{\partial H_T}{\partial p^i} \beta^i\right), \] 
\[\dot{\beta}^i \approx -\frac{\partial}{\partial q_i} \left( \frac{\partial H_T}{\partial q_i} \alpha_i + \frac{\partial H_T}{\partial p^i} \beta^i\right).\eqno(A9)\]
We now can equate the expressions for $\dot{\alpha}_i$ and $\dot{\beta}^i$ in eqs. (A7) and (A9) and then use eq. (A6) to eliminate $\alpha_i$ and $\beta^i$.
If the gauge generator is expanded
\[ G = \epsilon(t)G_0 + \dot{\epsilon}(t)G_1 + \ldots \epsilon^{(N)}(t)G_N\eqno(A10)\]
when there are $N + 1$ generations of constraints, then we find that 
\[\epsilon\left\{G_0, H_T\right\} + \dot{\epsilon}\left[ G_0 + \left\{G_1, H_T\right\}\right] + \ddot{\epsilon}\left[ G_1 + \left\{G_2, H_T\right\}\right]\nonumber\]
\[+ \ldots \epsilon^{(N)}\left[ G_{N-1} + \left\{G_N, H_T\right\}\right] + \epsilon^{(N+1)}\left[G_N\right] \approx 0.\eqno(A11)\]
Eq. (11) can be satisfied iteratively by taking
\[ G_N \approx \;({\rm{primary\;constraints}})\eqno(A12)\]
\[ G_{N-1} + \left\{G_N, H_T\right\}  \approx \;({\rm{primary\;constraints}})\nonumber\]
\[ \left\{G_0, H_T\right\} \approx \;({\rm{primary\;constraints}}).\nonumber\]
Only primary constraints appear in eq. (A12) as in  eq. (A7) the  weak inequality need only hold on the constraint surface on which the primary constraints vanish.


\begin{thebibliography}{99}
\bibitem{1} P.A.M. Dirac, \textit{Can J. Math.} \textbf{2}, 129 (1950).
\bibitem{2} P.A.M. Dirac, \textit{Lectures on Quantum Mechanics} (Dover, Mineola 2001).
\bibitem{3} M. Henneaux and C. Teitelboim,  \textit{Quantization of Gauge Systems} (Princeton U. Press, Princeton 1992).
\bibitem{4} D.M. Gitman and I.V. Tyutin, \textit{Quantization of Fields with Constraints} (Springer-Verlag, Berlin 1990).
\bibitem{5} K. Sundermeyer, \textit{Constrained Dynamics} (Springer-Verlag, Berlin, 1982).
\bibitem{6} A. Hanson, T. Regge and C. Teitelboim,  \textit{Constrained Hamiltonian Systems}, Acad. Naz. dei Lin. 1976.
\bibitem{7} N. Kiriushcheva and S.V. Kuzmin, \textit{Mod. Phys. Lett. A}, \textbf{21}, 899 (2006).
\bibitem{8} N. Kiriushcheva, S.V. Kuzmin and D.G.C. McKeon, \textit{Mod. Phys. Lett A} \textbf{20}, 1898 (2005).
\bibitem{9} N. Kiriushcheva, S.V. Kuzmin and D.G.C. McKeon, \textit{Mod. Phys. Lett A} \textbf{20}, 1961 (2005).
\bibitem{10} N. Kiriushcheva, S.V. Kuzmin and D.G.C. McKeon, \textit{Int. J. Mod. Phys.} \textbf{A21}, 3401 (2006).
\bibitem{11} N. Kiriushcheva and S.V. Kuzmin, \textit{Ann. Phys.}(N.Y.) \textbf{321}, 958 (2006).
\bibitem{12} R.N. Ghalati, D.G.C. McKeon and T.N. Sherry, \textit{Int. J. Mod. Phys.} \textbf{A22}, 4833 (2007).
\bibitem{13} D.G.C. McKeon, \textit{Class. Quant. Grav.} \textbf{23}, 3037 (2006).
\bibitem{14} U. Lindstrom and M. Rocek, \textit{Class. Quant. Grav.} \textbf{4}, 279 (1987).
\bibitem{15} J. Gegenberg, P.F. Kelly, R.B. Mann and D. Vincent, \textit{Phys. Rev.} \textbf{D37}, 3463 (1988).
\bibitem{16} L. Castellani, \textit{Ann. Phys.} (NY) \textbf{143}, 357 (1982).
\bibitem{17} J.M. Pons, D.C. Salisbury and L.C. Shepley, \textit{Phys. Rev.} \textbf{D55}, 658 (1997).
\bibitem{18} M. Henneaux, C. Teitelboim and J. Zanelli, \textit{Nucl. Phys. B} \textbf{332}, 169 (1990).
\bibitem{19} R. Banerjee, H.J. Rothe and K.D. Rothe, \textit{Phys. Lett.} \textbf{B462}, 248 (1999); ibid 479, 429 (2000). 
\bibitem{20} P.A.M. Dirac, \textit{General Theory of Relativity}, Ch. 26 (Princeton U. Press, Princeton 1996).
\bibitem{21} A.M. Polyakov,\textit{Phys. Lett.} \textbf{B103}, 207 (1981).
\bibitem{22} P.A.M. Dirac, \textit{Proc. R. Soc.} \textbf{A246}, 333 (1958).
\bibitem{23} C. Battle, J. Gomis, X. Gr$\grave{a}$cia and J.M. Pons, \textit{J. Math. Phys.} \textbf{30}, 1345 (1989).
\bibitem{24} S.V. Kuzmin and D.G.C. McKeon, \textit{Ann. Phys.} (N.Y.) \textbf{318}, 495 (2005).
\bibitem{25} M. Henneaux \textit{Principles of String Theory} (Plenum Press, New York 1988).
\bibitem{26} J. Gomis, M. Henneaux and J.M. Pons, \textit{Class. Quant. Grav.} \textbf{7}, 1089 (1990).
\bibitem{27} D.G.C. McKeon, \textit{Class. Quant. Grav.} \textbf{9}, 1495 (1992).
\bibitem{28} D. Salisbury, arXiv 0904.3993 (hist-ph).
\bibitem{29} C. Batlle, J. Gomis, J.M. Pons and N. Roman-Roy, \textit{J. Math. Phys.} \textbf{27}, 2953 (1986)
\bibitem{30} J. Labastide, M. Pernici and E. Witten, \textit{Nucl. Phys. B} \textbf{310},611 (1988).
\bibitem{31} D. Montano and J. Sonnenschein, \textit{Nucl. Phys. B.} \textbf{324}, 348 (1988).
\bibitem{32} N. Kiriushcheva and S.V. Kuzmin, \textit{Central Eur. J. Phys.} \textbf{9} 576 (2011).
\bibitem{33} A.M. Frolov, N. Kiriushcheva and S.V. Kuzmin, arXiv 0809.1198 (gr-qc).
\bibitem{34} J. Govaerts, \textit{Int. J.Mod. Phys. A}, \textbf{3625} (1990).
\bibitem{35} J. Govaerts,\textit{"Hamiltonian Quantization of Constrained Systems"} (Leuven University Press, 1991).
\bibitem{36} N. Kiriushcheva and S.V. Kuzmin, \textit{Eur. J. Phys. C}, \textbf{70}, 389, arXiv 0912.3396 (gr-qc).
\bibitem{37} D.G.C. McKeon,  \textit{Int. J. Mod. Phys. A} \textbf{A25}, 3453 (2010).


\vspace{.2cm}
\end{thebibliography}
\end{document}